\documentstyle[preprint,aps]{revtex}

\begin{document}
\draft
%
%-----------------------------------TEXT---------------------------
%
\title{Theory for the giant magneto-optical Kerr rotation in CeSb}
\author{U. Pustogowa, W. H\"ubner, and K. H. Bennemann}
\address{Institute for Theoretical Physics, Freie Universit\"at 
Berlin, Arnimallee 14, D-14195 Berlin, Germany}
\date{\today}
\maketitle
\begin{abstract}
We calculate the {\em linear} magneto-optical Kerr rotation angle for CeSb 
in the near-infrared spectral range. 
Using an exact formula for large Kerr rotation angles and a simplified 
electronic structure of CeSb we find at $\hbar \omega = $ 0.46~eV 
a Kerr rotation of 90$^{\circ}$ which then for decreasing $\omega $ jumps to 
-90$^{\circ }$ in very good agreement with recent experimental observations.
We identify the general origin of possible 90$^{\circ }$ polarization 
rotations from 
mainly optical properties and discuss its relation to the magnetic moments and  
magnetic dichroism of the material. 

\end{abstract}

\newpage

The search for large Kerr rotations in magneto-optics has been a longstanding 
subject of both basic and application-oriented research. 
The prediction and observation of giant Kerr 
rotations~\cite{pusto,Kirsche,prlKoop,Hue95} in the nonlinear magneto-optical 
Kerr effect (NOLIMOKE) on multilayer sandwiches and thin magnetic films have 
successfully demonstrated the enhanced sensitivity of nonlinear optics to 
magnetism, in particular on low-dimensional systems due to the reduction of 
symmetry. 
Although the detailed values of the enhanced Kerr rotation in NOLIMOKE depend 
on the electronic structure of the investigated system, the spin-orbit 
coupling strength, 
the magnetization direction in the sample and the polarization of the 
incoming light,  
the 90$^{\circ}$ nonlinear Kerr rotation in the longitudinal 
configuration and steep angle of incidence~\cite{Hue95}, is essentially 
independent of the frequency and results from symmetry considerations.  

Prior and also parallel to NOLIMOKE, the search for enhanced Kerr rotations 
has been pursued in the conventional linear magnet-optic Kerr effect 
(MOKE). 
Since the MOKE rotation of transition metals like Fe or Ni is typically in the 
range of only some 0.1$^{\circ}$ for optical frequencies one had to resort to 
particular magnetic rare-earth or uranium alloys with large spin-orbit 
coupling constants, large magnetic moments, and with special optical 
resonances in only one spin type to find relatively large Kerr rotations 
at low temperatures ($\approx $~1~K) and at lower frequencies. %($\ll$ 1 eV), 
Furthermore, the application of large external magnetic fields 
($\approx $~5~T) was necessary. 
In this way, a record-high MOKE rotation of 14$^{\circ}$ has been observed 
for CeSb in 1986 by Schoenes {\em et al.}~\cite{Reim86}. 
Very recently, however, the same group (Pittini {\em et al.}~\cite{moris}) 
observed the largest observable rotation of 90$^{\circ}$ and an abrupt jump of 
this rotation to -90$^{\circ }$ in CeSb by reducing the frequency from 
0.55 eV to 0.46 eV. 

In this paper, we present a simplified model for such a giant MOKE rotation. 
We argue that this large Kerr rotation occurs for a frequency set by both 
the plasma frequency and special optical transitions allowed only in one 
spin type, rather than the magnetic properties. 
Our explanation relies largely on an improved evaluation of the Kerr 
rotation for a model band structure which does not assume any small 
parameters in the off-diagonal components of the reflected electric field. 
We do not need all details of the complicated Ce monopnictides bandstructure 
to explain the large Kerr angle. 
In CeSb the effect of the Ce $f$ electrons is merely to 
spin-polarize and split up the Sb $p$ bands via hybridization. 
The simplified theory outlined in the following describes however already the 
physical origin of the large linear Kerr effect. 

The relation of MOKE signals (Kerr rotations or intensitiy measurements) to 
magnetic properties of the material is of special interest . 
Thus, MOKE is used for the determination of the direction and relative 
strength of the magnetization in saturated ferromagnetic materials. 
We show hereby analysing the reflectivity, the magnetic dichroism and the 
electrical-susceptibility-tensor elements that this special 90$^{\circ }$
Kerr-angle resonance is based on the behavior of the diagonal 
('nonmagnetic') susceptibility $\chi_{xx}(\omega )$ and thus is not  
related to the magnitude of the magnetization in the sample. 

To determine large Kerr rotation angles we have to use general 
expressions for the Kerr rotation and ellipticities in contrast to 
the usually taken linearized formulae which are valid if 
$\tan \varphi \approx \varphi $ holds. 
Following general ellipsometry arguments the complex polar Kerr angle 
$\kappa $ is defined as
% -------------- Equ. (1) {glw1}------------------------------------
	\begin{equation}
	\kappa \; = \; \frac {E_{y}} {E_{x}} \; = \; 
	\frac {\tan \varphi + i \tan \varepsilon }
		{1-i \tan \varphi \tan \varepsilon } \; ,
	\label{glw1}
	\end{equation}
% ------------------------------------------------------------
where, $\varphi $ is the Kerr rotation angle and  
$\varepsilon $ is the ellipticity. 
$E_{y}$ and $E_{x}$ are components of the reflected electric field. 
Solving Eq.~(\ref{glw1}) we find for the Kerr angle   
% -------------- Equ. (2) {loesung}------------------------------------
	\begin{equation}
	\varphi \; = \; \arctan \left( - \frac {1- |\kappa |^{2} } 
	{2 Re(\kappa ) } \pm 
	\sqrt {\frac {(1- |\kappa |^{2})^{2} } 
	{4 [Re(\kappa )]^{2} } + 1} \right) \; .
	\label{loesung}
	\end{equation} 
% ------------------------------------------------------------
The analysis of Eq.~(\ref{loesung}) yields that 180$^{\circ }$-jumps in 
$\varphi $ may occur when Re($\kappa $) is zero or infinity. 
The behavior of $\kappa $ results from 
the well known expression for the linear polar Kerr rotation 
% -------------- Equ. (3) {lin}------------------------------------
	\begin{equation}
	\kappa \;=\;-\frac{\chi _{xy}^{(1)}(\omega )}
	{\chi _{xx}^{(1)}(\omega )}
	\frac{1}{\sqrt {1 + \chi _{xx}^{(1)}(\omega )}} \; .
	\label{lin}
	\end{equation}
% ------------------------------------------------------------
Here, $\chi _{xx}^{(1)}$ and $\chi _{xy}^{(1)}$ denote the elements of the 
linear susceptibility tensor.
The complex value $\kappa $ describes the tangent of the angle, 
$\kappa = \tan \Phi_{K}$, with $\Phi_{K}=\varphi + i \varepsilon \,$. 
Note, for small Kerr rotations Eq.~(\ref{loesung}) reduces to 
$\varphi = Re(\kappa )$. 
Eq.~(\ref{loesung}) can be rewritten as 
% -------------- Equ. (4) {glphi0}------------------------------------
	\begin{equation} 
	\varphi \; = \; \frac {1} {2} \arctan \left( 
	\frac {2 Re(\kappa )} {1-|\kappa |^{2}}\right) + 
	\varphi_{0} \; ,
	\label{glphi0}
	\end{equation}
% ------------------------------------------------------------
with $\varphi_{0} = 0 $ for $ |\kappa |^{2} \le 1 $, 
$\varphi_{0} = 90^{\circ} $ for $ |\kappa |^{2} > 1,\; Re(\kappa ) \ge 0 $, 
and $\varphi_{0} = -90^{\circ} $ for $ |\kappa |^{2} > 1,\; Re(\kappa ) < 0$.
For details see Groot Koerkamp~\cite{Koerdipl,foot}. 
Thus, the behaviour of $\varphi $ for $Re(\kappa ) \rightarrow 0 $ depends on 
the value of $|\kappa |^{2}$. 
For $|\kappa |^{2} \le 1$, the case of small Kerr rotations, 
together with $Re(\kappa )$ also $\varphi $ goes to zero. 
In the other case, for $|\kappa |^{2} > 1$ , changing the sign of $Re(\kappa )$ 
yields to different values of $\varphi_{0} $, including a jump from 
$90^{\circ}$ to $-90^{\circ}$. 
So far, Kerr effect measurements realized only the first $|\kappa |^{2} \le 1$
case and no $\pm 90^{\circ}$ jump was observed. 
The analytical analysis of Eq.~(\ref{lin}) yields that $Re(\kappa )$ 
can get zero for $Re(\chi_{xx})=0$. 
Thus, we define two conditions for the occurence of a $\pm 90^{\circ }$ 
jump in $\varphi $
% -------------- Equ. (5) {condition}------------------------------------
	\begin{equation}
	|Im(\kappa )| \ge 1 \qquad \mbox{and} \qquad Re(\chi_{xx})=0
	\label{condition}
	\end{equation}

The ellipticity angle $\varepsilon$ is given by 
% -------------- Equ. (6) {epsi}------------------------------------
	\begin{equation}
	\varepsilon \; = \; \; \frac {1} {2} \arcsin \left( 
	\frac {2 Im(\kappa )} {1+|\kappa |^{2} } \right) \; .
	\label{epsi}
	\end{equation}
% ------------------------------------------------------------

Note, these results are of general interest, in particular the occurence of 
a "resonace" behaviour in the Kerr rotation. 

For describing now the infrared Kerr-spectrum of CeSb as measured by Pittini 
{\em et al.}~\cite{moris} we use {\em only one} feature of the electronic 
structure of CeSb close to the Fermi level,  
namely fairly flat and nearly parallel bands above and below the Fermi level 
$\varepsilon_{F}$ in the $\Gamma $ -- Z direction~\cite{Liecht}.
The $p$ states of Sb are split by approximately 0.6 eV. 
Furthermore, via hybridization with the $f$-electron minority-spin subband 
of CeSb (lying $\sim 3$ eV above $\varepsilon_{F}$) the Sb $p$ states 
have an induced spin polarization. 
Thus, there are favored transitions of one spin sort between the $p$ 
states of very high weight~\cite{dipol}. 
In this case, the interband parts of the diagonal and off-diagonal 
susceptibilities can be written as sums over transitions for one spin 
band only. 
Thus, 
% -------------- Equ. (7) ------------------------------------
	\begin{equation} 
	\chi_{xx} = \chi_{xx,intra} \; + \; 
	\sum_{l,l^{\prime}} L_{l,l^{\prime}} \; , 
	\; \; \; \; \; \; \; \; \;
	\chi_{xy} = -
	\frac {\lambda_{s.o.} } {\hbar \omega} 
	\sum_{l,l^{\prime}} L_{l,l^{\prime}} \; ,
	\end{equation}
% ------------------------------------------------------------
where $\chi_{xx, intra}$ denotes the intraband contribution important only 
in the diagonal elements of $\chi $,   
$L_{l,l^{\prime}}$ denotes the 
response from transitions between the bands $l$ and $l^{\prime }$ and 
$\lambda_{s.o.}$ is the spin-orbit coupling constant. 
Note, the factor $\lambda_{s.o.}/\hbar \omega $ results from including 
spin-orbit coupling to lowest order in the wave functions. 
However, this will not directly effect $\varphi $. 
The Kerr rotation is then calculated by using 
% -------------- Equ. (8) {kapl}------------------------------------
	\begin{equation}
	\kappa \;=\; \frac{\lambda_{s.o.}}
	{\hbar \omega } \; 
	\frac{1}{ \left( 1+ \frac {\chi_{xx,intra}} {L} \right) \; 
	\sqrt {1 + \chi _{xx}^{(1)}(\omega )}} \; .
	\label{kapl}
	\end{equation}
% ------------------------------------------------------------
For simplicity, we approximate the transitions between the (Sb) $p$ bands 
by a single atomic (nondispersive) Lorentzian
	\[
	\sum_{l,l^{\prime}} L_{l,l^{\prime}} \; = \; L \; := \; 
	\frac {1} {E_{f} - E_{i} - \hbar \omega + i \hbar \alpha } 
	\; ,
	\] 
with the band positions $E_{f}$ of the unoccupied final state and $E_{i}$ of 
the occupied initial state and the damping factor $\alpha $. 
Note, optical transitions between {\em minority}-spin electrons in 
these states strongly prevail.  
Furthermore, we describe intraband effects by a conventional Drude term 
	\begin{equation}
	\chi _{xx, intra} = - \frac {\omega_{pl}^{2}} 
	{ \omega (\omega + i\tau ) } \; ,
	\end{equation}
with the plasma frequency $\omega _{pl}$ and the damping $\tau $. 
Using these expressions one can then determine $\varphi $ with the help of 
Eq.~(\ref{glphi0}). 

For a further analysis and for comparison with experiment we also calculate 
the optical reflectivity R using 
	\begin{equation}
	R \; = \; \left| \frac {N-1} {N+1} \right| ^{2} \; ,
	\end{equation}
with the refraction index $N^{2} = \epsilon_{0} + i\epsilon_{1} = 
1 + \chi_{xx} + i\chi_{xy}$ .

We now present results for the Kerr rotation, the reflectivity and 
in particular the dependence of the large Kerr angle on the plasma 
frequency $\omega_{pl}$ and interband splitting $\Delta E = E_{f} - E_{i}$ 
referring to the dominant optical transition. 

In Fig. 1 the frequency dependence of the linear polar Kerr angle $\varphi $ 
and the ellipticity $\varepsilon $ are shown. 
Here $\varphi $ and $\varepsilon $ are calculated using Eqs.~(\ref{glphi0}), 
(\ref{epsi}), and (\ref{kapl}). 
Parameters are choosen such as to obtain the jump in $\varphi $ 
from -90$^{\circ }$ to +90$^{\circ }$ at precisely 0.46~eV as measured by 
Pittini {\em et al.}~\cite{moris}. 
Thus, we use for the interband splitting $\Delta E$ = 0.67~eV 
in agreement with the bandstructure of Liechtenstein 
{\em et al.}~\cite{Liecht} and for the damping $\alpha $ = 0.1~eV as usually 
taken for Kerr angle calculations. 
In the Drude term the parameters $\omega_{pl}$~=~0.93~eV 
and $\tau $ = 0.95~$\times $~10$^{-4}$~eV are used. 
The value for $\tau $ is taken from Kwon {\em et al.}~\cite{KwonS}. 
Note, the jump at 0.46 eV is really a jump from an angle -90$^{\circ }$ to 
an angle +90$^{\circ }$ without intermediate values. 
The ellipticity angle changes sign at 0.41~eV going from a minimum of 
-20$^{\circ }$ to the maximum of 35$^{\circ }$ . 
For comparison the experimental values by Pittini {\em et al.}~\cite{moris} 
are shown as dots. 

In Fig. 2 we show furthermore %for further comparison with experiment 
the frequency dependence of the optical reflectivity $R(\hbar \omega )$ 
in the energy range from 0 to 0.8~eV. 
For the Kerr angle spectrum the most important range of this curve  
is the deep minimum at 0.46~eV, at the same energy where the jump in 
$\varphi $ occurs. 
The discrepancy with respect to experimental results is presumably 
due to our simplified electronic structure of CeSb 
since we neglect interband transitions at large frequencies.  

Figs. 3 and 4 show the dependence of $\varphi (\hbar \omega )$ and, in 
particular, of the jump position on $\Delta E$ and the plasma frequency 
$\omega_{pl}$. 
Fig. 3 shows Kerr-angle spectra $\varphi(\hbar \omega)$ with plasma 
frequencies varying from 0.2~eV to 1.3~eV. 
Note, for plasma frequencies higher than a threshold value depending on 
$\Delta E$ the 180$^{\circ }$-jump vanishes, 
whereas for low plasma frequencies the jump amplitude is stable, while only 
the position of the jump moves to lower energies. 
The variation of the interband transition $\Delta E$ yields the 
opposite behavior, see Fig. 4. 
Here, $\Delta E$ changes from 0.2 eV to 1.6 eV. 
The jump in $\varphi $ vanishes for low $\Delta E$ and moves for 
large values of $\Delta E$ to higher energies. 
Thus, we find that the ratio of the plasma frequency $\omega_{pl}$ to 
$\Delta E$, which characterizes the interband transitions, 
is essential for the occurrence of a 180$^{\circ }$ jump in the Kerr angle. 

The position of the 180$^{\circ }$ jump in the Kerr angle $\varphi $ 
is determined by a zero of the real part of $\chi_{xx} $, in particular  
$\left( \chi_{xx,intra} + L \right) = 0$ . 
This equation yields a condition for the ratio of $\omega_{pl} / \Delta E$ 
at which the Kerr-angle jump may occur. 
Note, this equation reflects the ratio $\frac {\chi_{xx}} {\chi_{xy}}$ 
using two approximations, namely 
(i) $\chi_{xy} \sim \frac {\lambda_{s.o.}}{\hbar \omega }$ and 
(ii) assuming interband transitions only in 
one spin subband. 
Approximately, for $\omega_{pl} / \Delta E \le 1.5 $ the Kerr angle jump 
disappears. 
Moreover, there is no lower boundary for the occurrence of this 
jump as a function of the ratio $\omega_{pl} / \Delta E$~\cite{Bemerkung}. 

For all parameter sets we analyzed the realization of the jump 
conditions Eq.(~\ref{condition}). 
The second one, $Re(\chi_{xx})=0$ was fulfilled in every case indicating 
a resonace in $\varphi $. 
The additional analysis of the value of $|\kappa |$ or, at the resonance 
of $Im(\kappa )$, enable us to distinguish between less exciting 
and already discussed~\cite{Feil} resonances for $|Im(\kappa )| < 1$ 
and the 'degenerated' resonances with the $\pm 90^{\circ }$ jump in $\varphi $  
for $Im(\kappa ) \ge 1$.

Note, the jump in the Kerr rotation can be described also by a continuos 
rotation from 0$^{\circ }$ to 180$^{\circ }$, which are identical 
polarization angles. 
Then, the condition $Im(\kappa ) = 1$ indicates the boundary between 
those resonances, which return the angle to zero and in the other case those, 
which allow the continuos rotation. 
The jump occurs only as a result of implementing the usual boudary conditions: 
$\varphi =0$ Kerr rotations outside the 'excited' frequency region. 
 
In the energy range below $\approx $ 0.5 eV our results for the Kerr angle 
$\varphi (\hbar \omega)$ and Kerr ellipticity 
$\varepsilon (\hbar \omega )$ yield good agreement with the experimental 
data. 
The deviation for larger energies results from 
the neglect of further interband transitions. 
Note, we have included only the interband transition with $\Delta E$ in our 
model. 
For the same reason as already mentioned our calculation yields a too large 
reflectivity at 0.8 eV. 
The inclusion of more interband transitions, especially of transitions of the 
other spin, would decrease the reflectivity in this energy range and yield 
results for the Kerr rotation, ellipticity and reflectivity 
in better agreement with experiment for larger energies. 

Nevertheless, for the explanation of the Kerr angle jump it is not primarily 
important to know the exact electronic structure, but to fulfil the 
resonance condition combining the important features of the electronic 
structure with the plasma frequency. 
This concludes then the general model for large Kerr rotations due to the 
algebraic structure of $\kappa $ and $\varphi $, see 
Eqs.~(\ref{glw1},\ref{loesung},\ref{lin}). 

Historically, large linear Kerr rotations have been expected in materials 
with large spin-orbit coupling like in CeSb. 
In our model the spin-orbit coupling strength does of course influence the 
Kerr angle, namely a large spin-orbit coupling is necessary to realize 
$Im(\kappa ) \ge 1$, but not influences the jump position managed 
by $Re(\chi_{xx}) =0$. 
In addition, in the case of CeSb the interband splitting $\Delta E$ is 
affected by the spin-orbit coupling. 
Note, for the Kerr effect in transition metals the spin-orbit interaction does 
not appreciably influence the electronic bandstructure, but of course has to be 
included in the wave functions. 

In view of our model, related materials with pronounced interband 
transitions due to flat $f$ and $d$ bands could exhibit similar behavior. 
However, due to the resonance character of the $\varphi $ enhancement 
and the Kerr-angle jump, slight changes of the parameters and dominant optical 
transitions might already suppress the giant Kerr effect~\cite{Bebem}. 
This may explain why for example CeBi with similar $\lambda_{s.o.} $ and 
magnetic moments exhibits only a Kerr angle of about 
-9$^{\circ }$ ~\cite{Pitti2}.
The different shapes of $\varphi (\omega )$ for CeSb and CeBi might suggest 
already differences in the dominant dipole transitions and that in CeBi 
in contrast to CeSb the resonance does not occur (see $\varphi \simeq 
-9^{\circ } \longrightarrow \varphi \sim 3^{\circ } $ for $\hbar \omega $ 
= 0.35 eV to 0.5 eV). 

Mostly interesting is the relation of the discussed Kerr-rotation jump 
to the magnetic properties of the material. 
In view of this we calculate the magnetic dichroism in MOKE. 
With reference to the magnetic dichroism in circularly polarized light 
we define the dichroism for linearly incident polarization 
	\begin{equation}
	d \; = \; \frac {I(+M)-I(-M)} {I(+M)+I(-M)} \; ,
	\end{equation}
with $I(\pm M)$ the intensities for inversed magnetization directions. 
Using for the intensities of the reflected light 
$I=|E_{refl}|^{2} = |\chi \times E_{incident}|^{2} $ we find 
$I(\pm M) \sim |\chi_{xx} \pm i \chi_{xy}|^{2}$ and the dichroism
	\begin{equation}
	d \; = \; 2 
	\frac { Im(\chi_{xx}) Re(\chi_{xy}) - Re(\chi_{xx}) Im(\chi_{xy}) } 
	{|\chi_{xx}|^{2} + |\chi_{xy}|^{2}} \; .
	\end{equation}

In Fig. 5 we illustrate the different frequency dependences of the 
magnetic dichroism $d$ and the absorptive part of the magnetic susceptibility 
tensor element $Im(\chi_{xy})$. 
We find a strong dichroism of more than 70\% at the $\varphi $-jump frequency 
whereas the spin-polarized absorption depends only on the electronic 
structure. 
The maximum in $Im(\chi_{xy})$ occurs at $\Delta E =0.67$ eV. 
In previous more detailed microscopic calculations of $\chi_{xy}$~\cite{sursc} 
we found a linear dependence of amplitudes of pronounced maxima in $\chi_{xy}$ 
on the magnetic moment and this seems to be a general property. 
On the other hand, inspecting Eq.(~\ref{lin}) (small Kerr effect) 
for a weak varying denominator we find $\varphi \sim \chi_{xy} \sim M$. 
For the $\varphi $-jump case, in contrast, the Kerr rotation and the dichroism 
are determined by nonmagnetic optical properties ($Re(\chi_{xx})=0$) not 
resolving the magnetism of the material. 
Thus, we analyzed a case of polarization rotation in a magnetic material 
not caused by magnetic properties and, strictly speaking, cannot register this 
effect as magneto-optical Kerr effect. 
Experimentally, the usual MOKE and the here discussed polarization rotation 
can be distinguished by an additional analysis of the optical 
reflectivity.  

It would be of interest to determine also the {\em nonlinear} Kerr rotation 
in CeSb. 
Generally, the nonlinear Kerr effect involving more transitions will be a 
more sensitive probe of the electronic structure. 
Also, the nonlinear Kerr rotation would illustrate the different nature of 
giant Kerr rotations in linear and nonlinear optics. 
The theoretical analysis for this follows from previous 
studies~\cite{pusto,Hue95}. 

We thank J. Schoenes and P. Wachter for providing us with 
unpublished data and for stimulating discussions.  

\newpage
\newpage
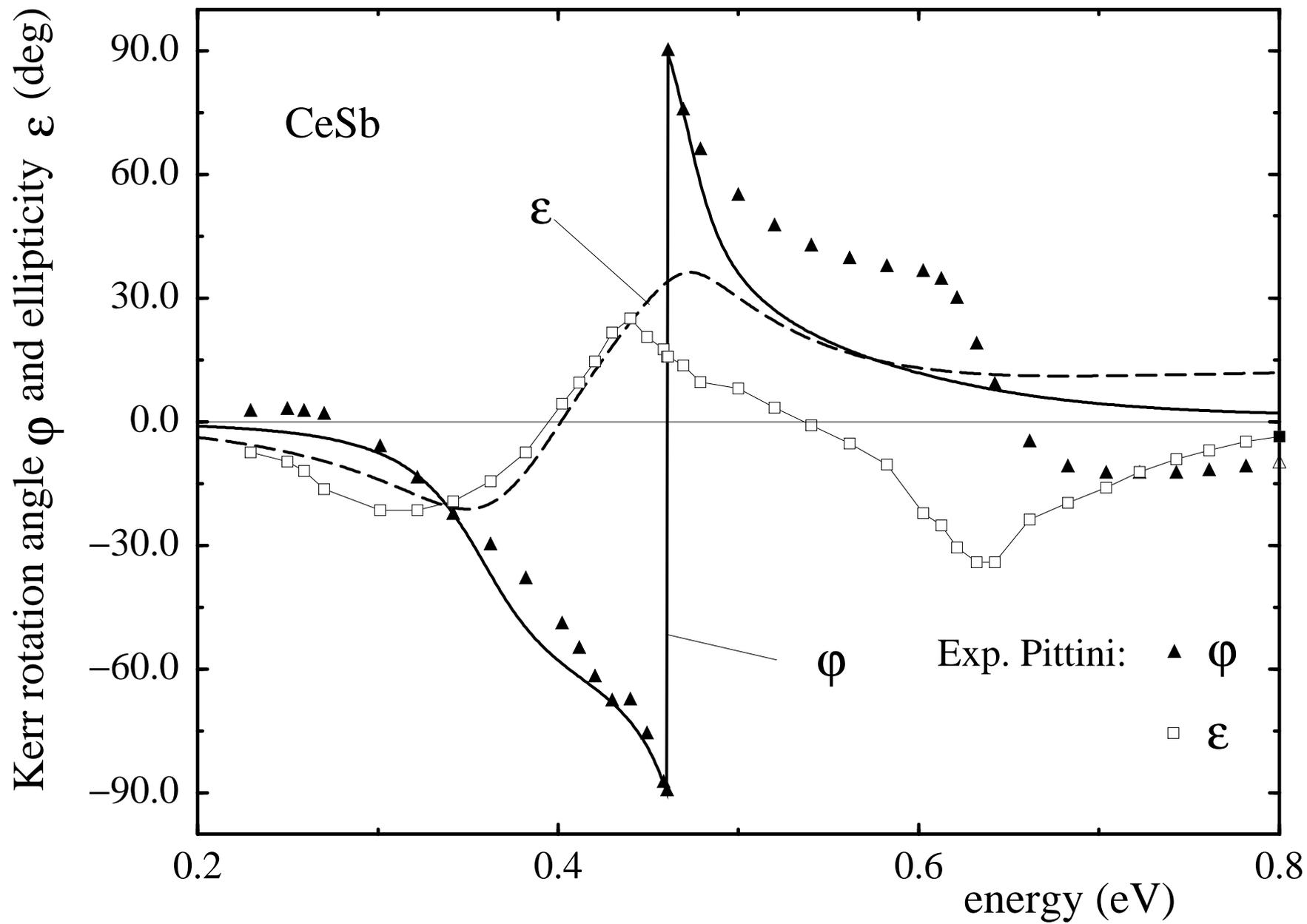
\begin{figure}
\caption[]{Calculated frequency dependence of the linear magneto-optical 
Kerr-angle $\varphi$ (solid curve) and the ellipticity angle $\varepsilon$ 
(dashed curve) of CeSb calculated with a Lorentz-type susceptibility.
For comparison also experimental results of Pittini {\em et al.} for 
$\varphi (\hbar \omega)$ and $\varepsilon (\hbar \omega )$ are shown.}
\end{figure}
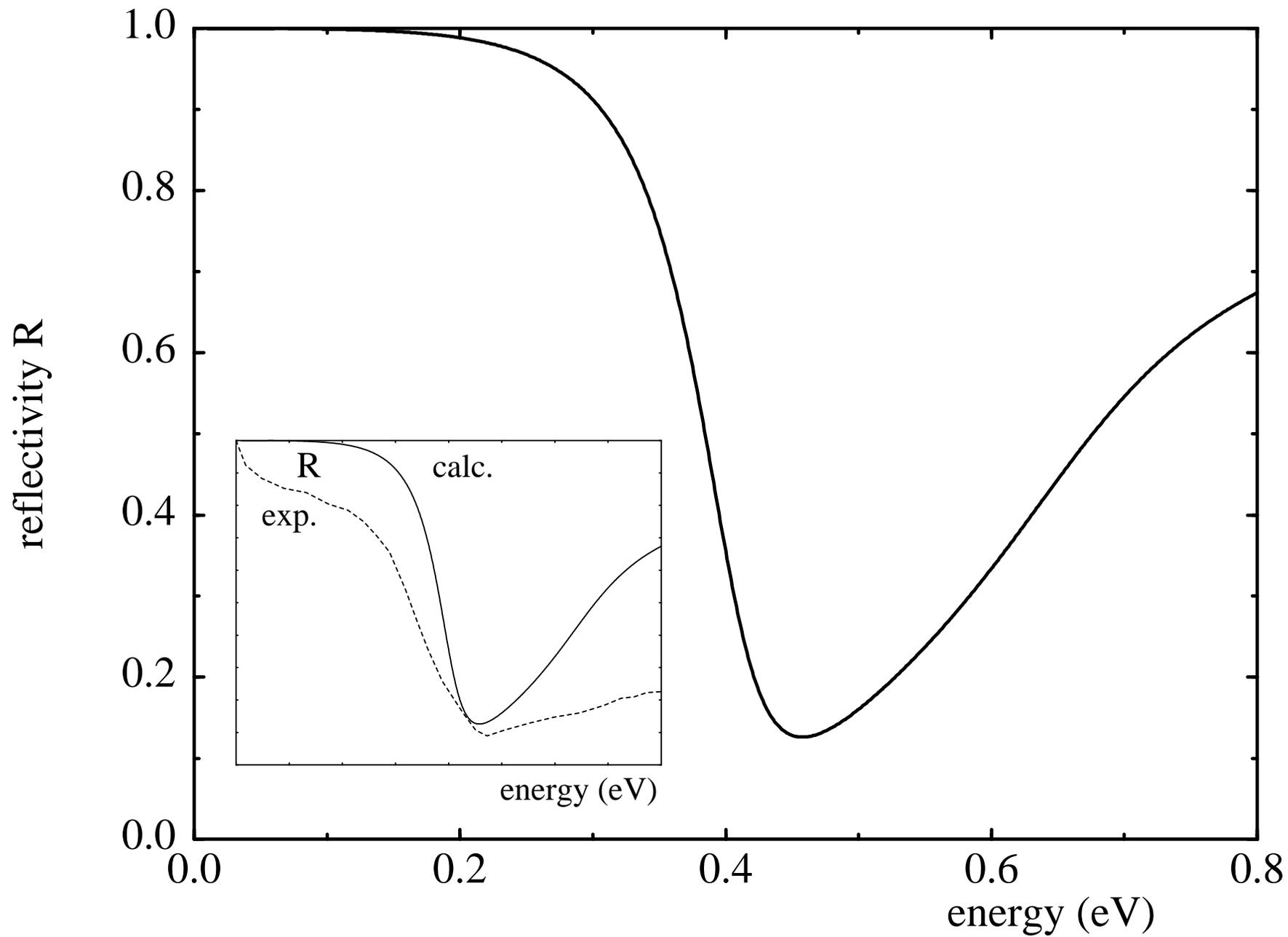
\begin{figure}
\caption[]{Optical reflectivity $R$ of CeSb calculated from the diagonal 
susceptibility $\chi_{xx}(\omega)$. Experimental values are also shown.}
\end{figure}
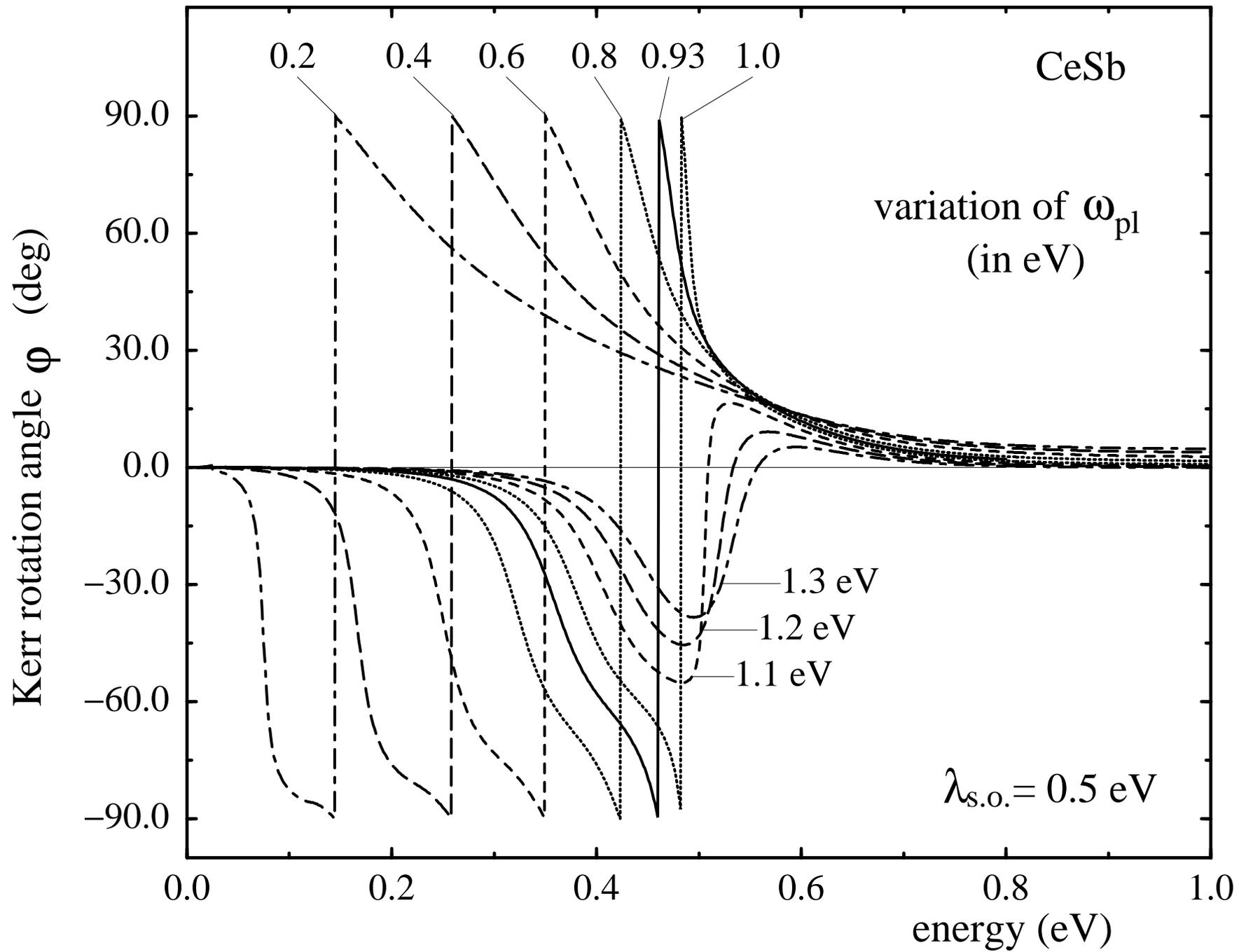
\begin{figure}
\caption[]{Dependence of the Kerr angle spectra $\varphi (\hbar \omega)$ 
of CeSb on the value of the plasma frequency $\omega _{pl}$. 
For every curve the value of $\omega_{pl}$ (in eV) is shown.}
\end{figure}
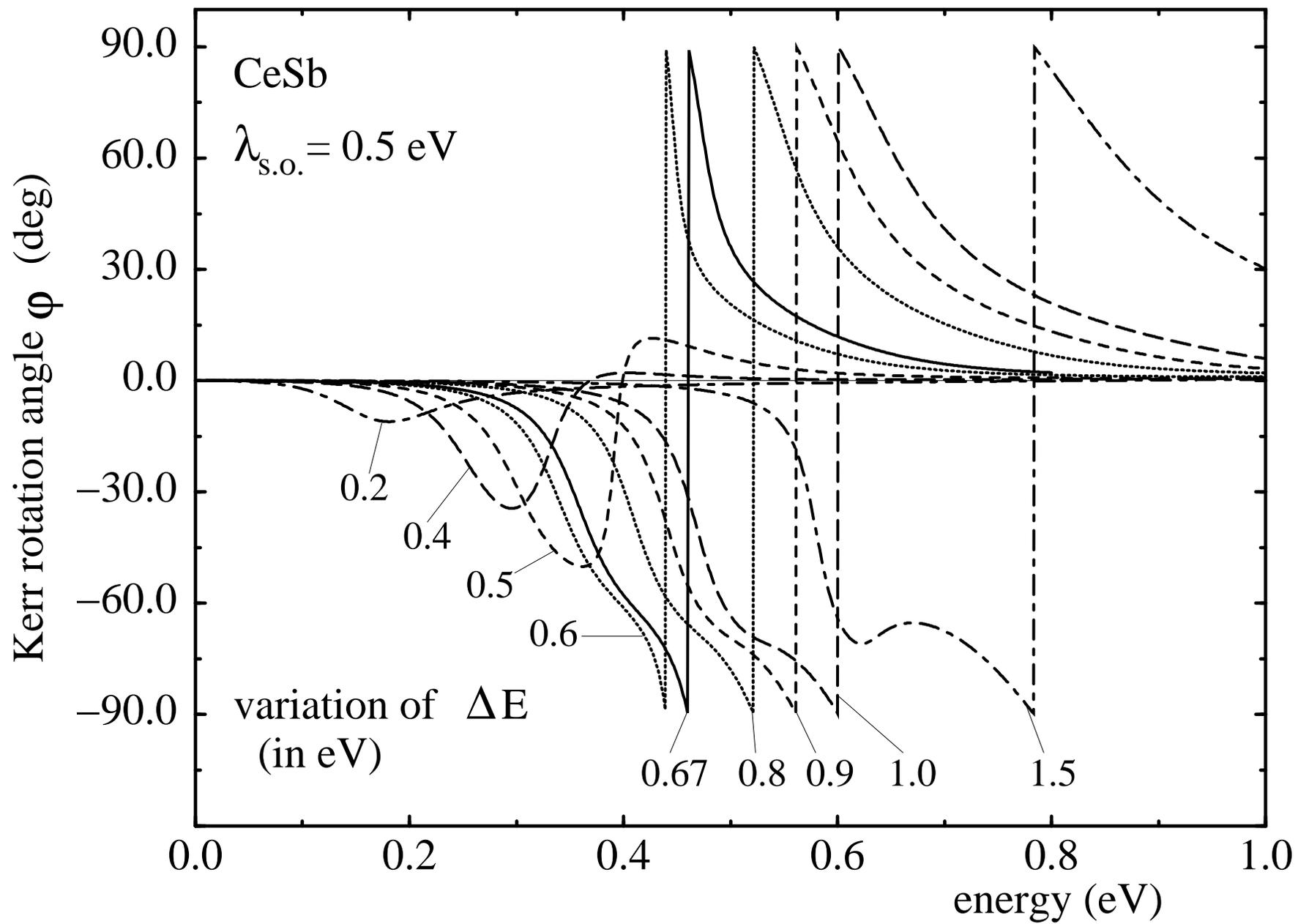
\begin{figure}
\caption[]{Dependence of the Kerr angle spectra $\varphi (\hbar \omega)$ 
on the interband transition energy $\Delta E$. 
For every curve the corresponding value of $\Delta E$ (in eV) is shown .}
\end{figure}
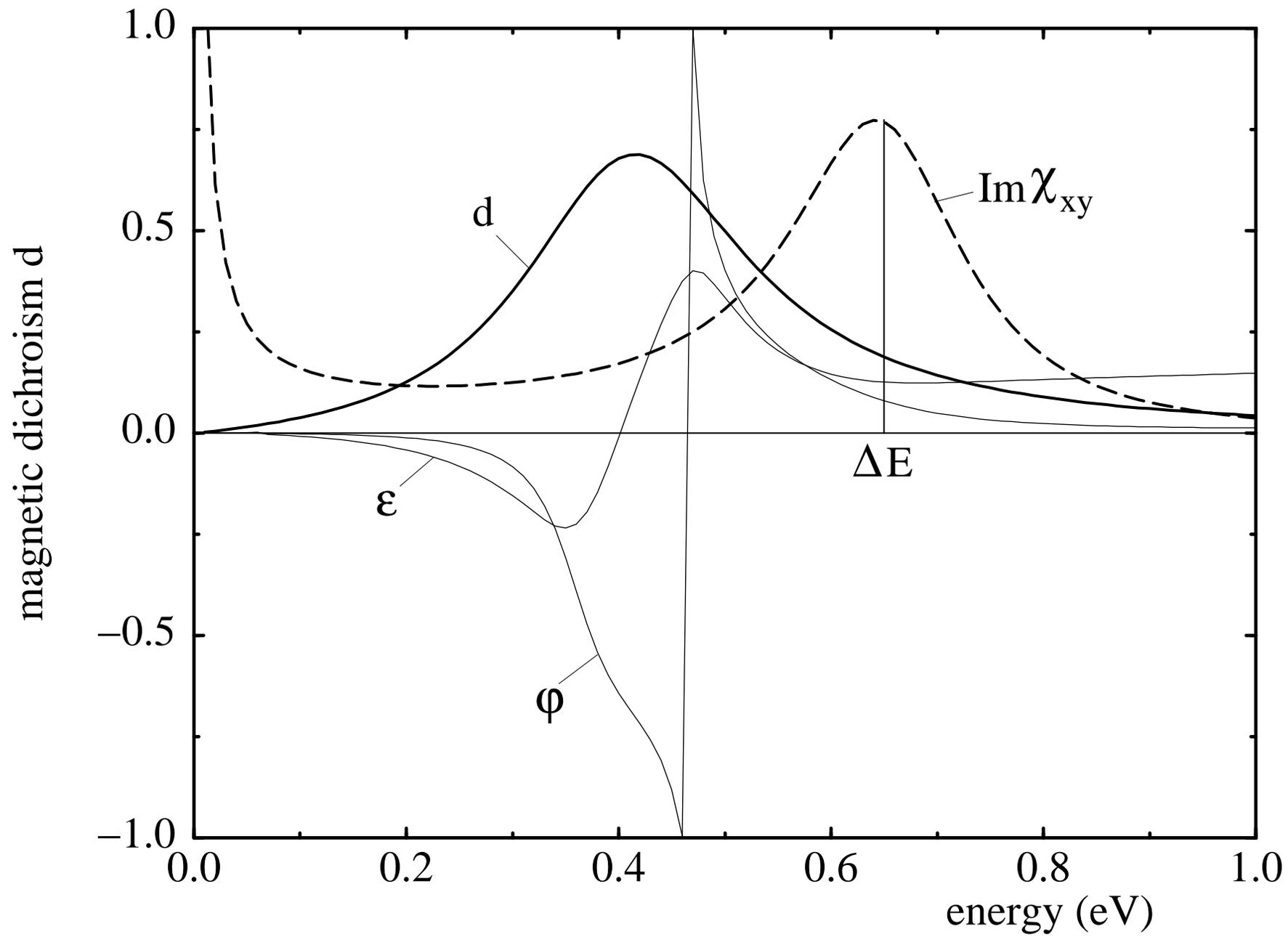
\begin{figure}
\caption[]{Frequency dependence of the magnetic dichroism $d$ and the 
absorptive part of the magnetic susceptibility $Im(\chi_{xy})$.}
\end{figure}

\begin{references}
\bibitem[1]
{pusto} U. Pustogowa, W. H\"ubner, and K. H. Bennemann, 
Phys. Rev. B {\bf 49}, 10031 (1994).
\bibitem[2]
{Kirsche} H. A. Wierenga, W. de Jong, M. W. J. Prins, Th. Rasing,
R. Vollmer, A. Kirilyuk, H. Schwalbe, and J. Kirschner, Phys. Rev. Lett.
{\bf 74}, 1462 (1995).
\bibitem[3]
{prlKoop} B. Koopmans, M. Groot Koerkamp, Th. Rasing, and H. van den Berg,
Phys. Rev. Lett {\bf 74}, 3692 (1995).
\bibitem[3]
{Hue95} W. H\"ubner and K. H. Bennemann, Phys. Rev. B. {\bf 52}, 13411
(1995).
\bibitem[5]
{Reim86} W. Reim, J. Schoenes, F. Hulliger, and O. Vogt, J. Magn. Magn. 
Mater. {\bf 54-57}, 1401 (1986).
\bibitem[6]
{moris} R. Pittini, J. Schoenes, O. Vogt, and P. Wachter, 
Phys. Rev. Lett. {\bf 77}, 944 (1996). 
\bibitem[7]
{Koerdipl} M. Groot Koerkamp, Diploma thesis, University of Nijmegen (1995).
\bibitem[8]
{foot} An equivalent solution of Eq.~(\ref{glw1}) is given by 
% -------------- Equ. (3) {phi}------------------------------------
Groot Koerkamp~\cite{Koerdipl} with
	\[
	\varphi \; = \; \frac {1} {2} \; \arctan \left( 
	\frac {2 Re(\kappa )} {1-|\kappa |^{2} } \right) \; .
	\]
% ------------------------------------------------------------
The solution of Eq. (1) for $\varphi $ and $\varepsilon $ starts with 
rewriting this complex equation into two real equations for $Re(\kappa )$ 
and $Im(\kappa )$. 
Whereas Eq. (2) was found by straight-forward substitution of $\tan 
\varepsilon $, 
another solution is found adding $|Re(\kappa )|^{2}$ and $|Im(\kappa )|^{2}$ 
and subsequent using the addition theorem between 
$\tan \varphi $ and $\tan 2\varphi $. 
Than, the following inversion of $\tan 2\varphi $ yields to an unphysical 
halving of the definition range to $[-\pi /4,\pi /4]$, which is compensated  
by the angle $\varphi_{0} $ extending the definition range to 
$[-\pi /2 , \pi /2]$. 
\bibitem[9]
{Liecht} A. I. Liechtenstein, V. P. Antropov, and B. N. Harmon, Phys. Rev. 
B {\bf 49}, 10770 (1994).
\bibitem[10]
{dipol} Note, in magneto-optics the observance of $\Delta l = \pm 1$ for dipole 
transitions is not required following from the addition of spin-orbit 
interaction. Furthermore, due to hybridization the discussed bands contain 
beside the $p$ character also some $f$-electron character. 
\bibitem[11]
{KwonS} Y. S. Kwon, T. Suzuki, and T. Kasuya, 
J. Magn. Magn. Mater. {\bf 116}, 73 (1992). 
\bibitem[12]
{Bemerkung} Note, these conditions change when transitions in both majority 
and minority spin bands have to be considered. 
\bibitem[13]
{Feil} H. Feil and C. Haas, Phys. Rev. Lett. {\bf 58}, 65 (1987).
\bibitem[14]
{Bebem} Note the differences between CeSb and CeBi regarding the ellipticity 
and the off-diagonal conductivity $\sigma_{xy}$, s. Pittini {\em et al.}. 
Note, CeBi exhibits less structure in $\sigma_{1xy}$ and $\sigma_{2xy}$ 
and a different frequency dependence suggesting different role of important 
electric dipole transitions. 
\bibitem[15] 
{Pitti2} R. Pittini, J. Schoenes, and P. Wachter, to be published (1996).
\bibitem[16]
{sursc} U. Pustogowa, W. H\"ubner, K. H. Bennemann, Surf. Sc. {\bf 307-309}, 
1129 (1994).
%
\end{references}
\end{document}